\newcommand{\nd}{Ni(C$_5$D$_{14}$N$_2$)$_2$N$_3$(PF$_6$)}

\documentstyle[prl,aps,twocolumn]{revtex}
\begin{document}
\input{psfig.sty}
\draft

\twocolumn[\hsize\textwidth\columnwidth\hsize\csname
@twocolumnfalse\endcsname

\title{Field-induced 3- and 2-dimensional freezing in a quantum spin liquid.}

\author{Y. Chen,$^\dag$, Z. Honda$^{\ddag}$, A. Zheludev,$^\sharp$,  C. Broholm,$^{\dag,\ast}$,
K. Katsumata$^{\ddag}$ and S. M. Shapiro$^{\sharp}$.}
\address{ $\dag$ Department of Physics and
Astronomy, Johns Hopkins University, Baltimore, MD 21218, USA.\\
$\ddag$ RIKEN (The Institute of Physical and Chemical Research),
Wako, Saitama 351-0198, Japan.\\ $\sharp$ Physics Department,
Brookhaven National Laboratory, Upton, NY 11973-5000, USA.\\
$\ast$ NIST Center for Neutron Research, National Institute of
Standards and Technology, Gaithersburg, MD 20899, USA.}

\date{\today}
\maketitle
\begin{abstract}
Field-induced commensurate transverse magnetic ordering is
observed in the Haldane-gap compound \nd\ by means of neutron
diffraction. Depending on the direction of applied field, the
high-field phase is shown to be either a 3-dimensional ordered
N\'{e}el state or a short-range ordered state with dominant
2-dimensional spin correlations. The structure of the high-field
phase is determined, and properties of the observed quantum phase
transition are discussed.
\end{abstract}

\pacs{} ]

The one-dimensional (1D) integer-spin Heisenberg antiferromagnet
(AF) demonstrates unique quantum-mechanical behavior that is
inconsistent with the conventional semi-classical model of
magnetism. Due to zero-point quantum spin fluctuations, that
destroy long-range order (LRO) even at $T=0$, the ground state is
a spin-singlet with short-range (exponentially decaying) spatial
spin correlations, and is often referred to as a ``quantum spin
liquid''. The main feature of the magnetic excitation spectrum is
the so-called Haldane energy gap \cite{Haldane83}. Quite
remarkable is the behavior of 1D and quasi-1D integer-spin AFs in
applied magnetic fields. The effect of the field is to suppress
zero-point fluctuations, and restore a gapless spectrum. The
result is a quantum phase transition at a certain critical field
$H_c$, to a N\'{e}el-like state with long-range order that may be
characterized as a ``spin solid''. Thus, in an unusual twist, a
{\it uniform} field induces {\it staggered} magnetization in
quantum-disordered spin chains.

\begin{figure}
\psfig{file=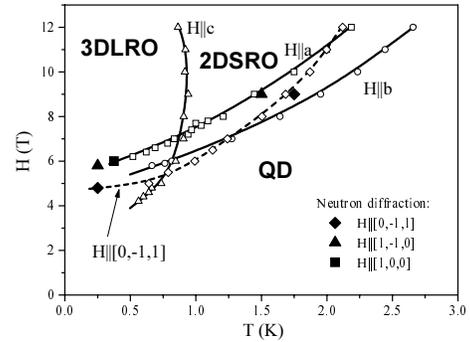,width=3.0in,angle=0}\caption{Field-temperature
phase diagram of NDMAP. A quantum-disordered (QD) phase with a gap
is seen in low fields. The high-field phase is characterized by
static long-range order (3D-LRO) or quasi-2D short-range order
(2D-SRO). Open symbols from specific heat measurements, as in
Ref.~\protect\cite{Honda99}. Solid symbols from neutron
diffraction.} \label{phase}
\end{figure}

The magnetization process of a $S=1$ 1D AF is now rather well
understood theoretically
\cite{AffleckH90,AffleckH91,Sakai93,Golinelli93,Sakai94,Mitra94}.
For the isotropic case the problem was shown to be equivalent to
Bose condensation in one dimension \cite{AffleckH91,Nicopoulus91}.
Many theoretical results were confirmed in experimental studies of
real quasi-1D compounds. For a long time however, the actual phase
transition remained inaccessible for experimental investigation.
In some materials (such as Y$_2$BaNiO$_5$, for
example\cite{YBANO}), the value of $H_c$ is prohibitively high. In
other compounds ({\it e.g.}, NENP \cite{NENP}), the transition
does not occur due to certain structural features, and is instead
replaced by a broad cross-over
phenomenon\cite{Chiba91,Kobayashi92,Enderle00}. A true phase
transition at $H_c$ has been experimentally observed only
recently, in the quasi-1D $S=1$ AF materials NDMAP
\cite{Honda98,Honda99,Honda00} and NDMAZ \cite{Honda97}. Specific
heat, magnetization and ESR studies have provided a comprehensive
picture of the $H-T$ phase diagram \cite{Honda98,Honda99,Honda00},
a refined version of which is shown in Fig.~\ref{phase}.
Nevertheless, to date, the nature of the high-field phase remained
undetermined, and no direct evidence of staggered LRO has been
obtained. In particular, preliminary neutron scattering
experiments of Ref. \cite{fax} failed to detect any magnetic Bragg
peaks above $H_c$. In the present work we report a high-field
magnetic neutron diffraction study of NDMAP. We find that the
high-field phase is characterized by a commensurate ordering of
spin components perpendicular to the field direction.
Surprisingly, the nature of the high-field phase is extremely
sensitive to direction of applied field: depending on the
experimental geometry, for $H>H_c$ the system either develops true
3D LRO, or static quasi-2-dimensional short range order.

Magnetic properties of NDMAP are due to $S=1$ Ni$^{2+}$ spins that
are assembled into AF chains running along the $c$ axis of the
orthorhombic $Pnmn$ crystal structure ($a=18.05$~\AA,
$b=8.71$~\AA, $c=6.14$~\AA). Haldane gap excitations and magnetic
interactions in this system were previously investigated by means
of zero-field inelastic neutron scattering \cite{Zheludev00}.
In-chain AF interactions are dominant and correspond to a
Heisenberg exchange constant $J=2.8$~meV. The triplet of Haldane
excitations is split by easy-$(a,b)$ plane anisotropy, and the gap
energies are $\Delta_z=1.9$~meV and $\Delta_{\bot}\approx
0.47$~meV, for excitations polarized along and perpendicular to
the anisotropy axis, respectively. The coupling between chains
along the $b$ axis is rather weak, $J_y=2\cdot 10^{-3}$~meV, and
barely detectable along $a$, $|J_x|<5\cdot 10^{-4}$~meV.

The present neutron diffraction experiments were performed at the
SPINS cold-neutron 3-axis spectrometer at the National Institute
of Standards and Technology Center for Neutron Research. A 150~mg
sample was mounted with either the $(0,k,l)$, $(h,h,l)$ or
$(h,k,k)$ reciprocal-space planes in the scattering plane of the
spectrometer. Most measurements were performed using $(^{58}{\rm
Ni guide})-80'-80'-240'$ or $(^{58}{\rm Ni guide})-40'-40'-240'$
collimations, with a Be-filter positioned in front of the sample
to eliminate higher-order beam contamination, and $E_f=5$~meV or
$E_f=3$~meV fixed final energy neutrons. Sample environment in all
cases was a pumped He-3 cryogenic insert and a 9~T or 7~T
superconducting magnet, with the field applied vertically, {\it
i.e.}, perpendicular to the scattering plane.

\begin{figure}
\psfig{file=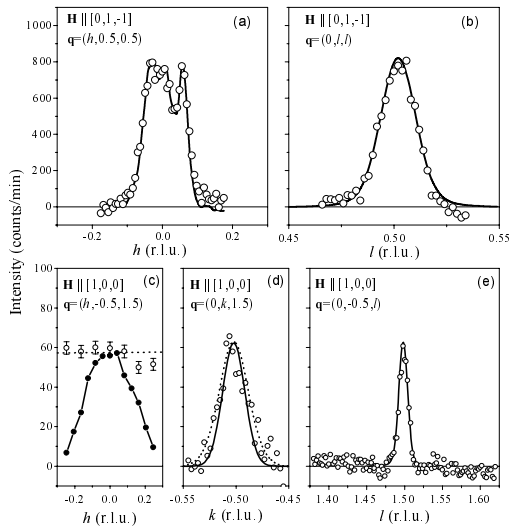,width=4.5in,angle=0}\caption{Typical elastic
scans (open symbols), showing magnetic Bragg scattering in NDMAP
at $T=0.25$~K, $H=9$~T$||[0,-1,1]$ (a,b) and $T=0.375$~K,
$H=7$~T$||[1,0,0]$ (c-e). The solid lines and symbols represent
experimental wave vector resolution. Dashed lines as described in
the text.}\label{exdata1}
\end{figure}

True 3D commensurate AF LRO was observed in a crystal mounted in
the $(h,k,k)$ zone, with the magnetic field applied along the
$[0,-1,1]$ direction. At $T=0.25$~K the transition occurs at
$H_c=4.8$~T, and manifests itself in the appearance of new Bragg
reflections, at $(h,\case{2n+1}{2},\case{2n+1}{2})$ ($h$-even
integer; $n$- integer) reciprocal-space positions. No additional
peaks were found at odd-$h$ reciprocal-space points.
Figures~\ref{exdata1}(a) and \ref{exdata1}(b) show typical scans
through the $(0,0.5,0.5)$ peak measured at $H=9$~T applied field
(symbols). The peaks are resolution-limited along both the
$(1,0,0)$ and $(0,1,1)$ directions. Wave vector resolution
(Fig.~\ref{exdata1}, solid lines) was either calculated or
directly measured in our experiment. The multiple peaks seen in
the rocking scan in Fig.~\ref{exdata1}(a) result from sample
mosaic. Geometrical limitations imposed by the use of the bulky
9~T magnet prevented us from measuring the width of the peak in
the vertical $[0,1,-1]$ direction. However, the substantial values
of the $b$ and $c$ axis coupling constants, as well as the
resolution-limited peak widths along $a$, where magnetic
interactions are weakest, suggest that the observed reflections
indeed have zero intrinsic width in all three directions, and
represent true 3D long-range order. This is also consistent with
the measured absolute values of magnetic intensities, as discussed
below.

To determine the spin arrangement in the high-field ordered phase,
intensities of (0,0.5,0.5) (2,0.5,0.5) (4,0.5,0.5) (2,1.5,1.5) and
(4,1.5,1.5) magnetic Bragg peaks in the $(h,k,k)$ plane were
measured in rocking scans at $T=0.25$ and $H=9$~T.  Magnetic
intensities were brought to the absolute scale by comparing them
to nuclear intensities measured in the same configuration at room
temperature (the exact low-temperature structure of NDMAP has not
been determined to date). The data were analyzed using a simple
collinear model. As the energy scale of in-chain AF interactions
is much larger than that defined by the applied magnetic field (
$J\gg g\mu_{\rm B}H$), the ordered moment is expected to be
perpendicular to the field direction. The $(a,b)$-easy plane
anisotropy ensures that it should also be perpendicular to the
chain-axis.  For a field applied along $[0,-1,1]$ the spins are
thus along $(1,0,0)$, as confirmed by recent ESR
experiments\cite{Honda99}. This allows only one magnetic structure
that would be consistent with the observed selection rule for
magnetic Bragg peaks. Varying only the sublattice magnetization
$m$, we obtained a good fit to the data ($\chi^2=1.6$) with
$m=1.13(5)$~$\mu_{\rm B}$. This value is to be compared to
$m=2$~$\mu_{\rm B}$ for a fully saturated $S=1$ antiferromagnet.

\begin{figure}
\psfig{file=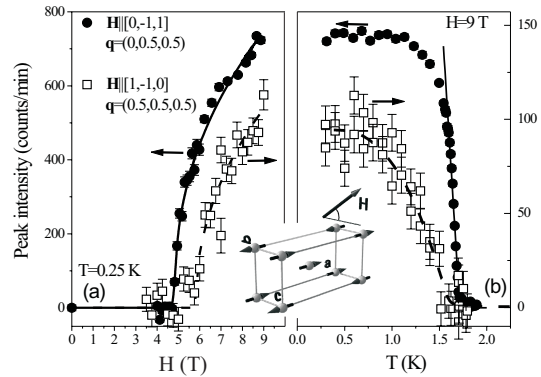,width=3.2in,angle=0}\caption{Measured field
(a) and temperature (b) dependence of the ordered moment in NDMAP
for a magnetic field applied in the $(b,c)$ plane. Solid lines are
power-law fits to the data, as described in the text. Inset: A
schematic view of the spin arrangement in the high-field
phase.}\label{vst}
\end{figure}

The measured field dependence of the $(0,0.5,0.5)$ magnetic Bragg
peak intensity at $T=0.25$~K is shown in Fig.~\ref{vst}(a) (solid
circles), and is consistent with our expectations for an
easy-plane Haldane system at $T\rightarrow 0$. The data were
analyzed assuming a power-law field dependencies of the Bragg
intensity and ordered moment: $ I(H)\propto (H-H_c)^{2\beta}$. A
fit of this equation to the data (Fig.~\ref{vst}(a), solid line)
yields $H_c=4.81(1)$~T and $\beta=0.207(5)$. Since the transition
at $H_c$ occurs through a softening of the lowest-energy Haldane
gap excitation \cite{AffleckH90}, $H_c$ is determined by the
Haldane gap energies. It is straightforward to show that for a
magnetic field applied at angle $\alpha$ to the anisotropy axis
$c$ is given by: $\mu_B H_c=\Delta_z\Delta_{\bot}/\sqrt
{g_z^2\Delta_z^2\cos^2\alpha+g_{\bot}^2\Delta_z\Delta_{\bot}\sin^2\alpha}$.
In this formula $g_z=2.1$ and $g_{\bot}=2.17$ are components of
the Ni$^{2+}$ gyromagnetic tensor along and perpendicular to the
anisotropy axis, respectively \cite{Honda98}. For
$\bbox{H}||[0,-1,1]$, $\alpha=54.8^{\circ}$ and this equation
gives $H_c=5.4$~T, in reasonable agreement with our experimental
result. In the case of broken axial symmetry, for a {\it single}
chain in a magnetic field at $T=0$, the transition is expected to
fall in the 2D Ising universality class, with magnetic field
taking the role of effective temperature\cite{AffleckH91}. The
order parameter critical exponent for this model is $\beta=0.125$.
For NDMAP however, inter-chain interactions along the $b$ axis can
be considered substantial, and 3D Ising behavior, with
$\beta\approx 0.31$, may be expected. The measured critical
exponent falls in between these two values and is characteristic
of a dimensional crossover regime.

The temperature dependence of the $(0,0.5,0.5)$ magnetic Bragg
intensity measured at $H=9$~T is shown in Fig.~\ref{vst}(b) (solid
circles) . An analysis of the data assuming a power law (solid
line) yields $T_c=1.7(1)$~K. The critical exponent was found to be
indistinguishable from the mean field value $\beta=0.5$,
indicating that at $T\approx 1.5$~K the true critical region is
too narrow to be investigated in the present experiment.

A striking result of this work is that for a magnetic field
applied perpendicular to the chain-axis, the high-field phase is
no longer a 3D-ordered state, but has predominantly 2-dimensional
spin correlations. For the crystal mounted in the $(0,k,l)$ zone,
at $T=0.38$~K and $\bbox{H}||[1,0,0]$, magnetic scattering was
detected at $(0,\case{2n+1}{2},\case{2m+1}{2})$ ($n$, $m$-integer)
reciprocal-space points above $H_c=6$~T. $k$- and $l$-scans
through the $(0,-0.5,1.5)$ reflections contain well-defined peaks,
as shown in Figs.~\ref{exdata1}(d) and ~\ref{exdata1}(e)
(symbols). Scans perpendicular to the $(b,c)$ plane however,
reveal that the scattering is concentrated in Bragg {\it rods}
along the $(1,0,0)$ direction, rather than Bragg {\it peaks}, as
in the $\bbox{H}||[0,-1,1]$ case. Figure.~\ref{exdata1}(c) shows
an $h$-scan through the $(0,-0.5,1.5)$ position, that could only
be performed in a limited range due to geometrical constraints.
The measured intensity is independent of $h$. Note that the
measured vertical resolution (Fig.~\ref{exdata1}(c), solid line),
is sufficient to observe a well-defined Bragg peak. The rod nature
of magnetic scattering implies that static spin correlations along
the $a$ axis are absent in the system, and that magnetic ordering
occurs only within individual $(b,c)$ planes. This is consistent
with the $a$-axis being the direction of weakest magnetic
interactions. Similar behavior was observed for a magnetic field
applied along the $[1,-1,0]$ direction, where magnetic scattering
was observed above $H_c=5.8(2)$~T, at $T=0.25$~K, and also found
to be concentrated in Bragg rods parallel to the $a$ axis. In this
geometry, the rods cross the $(h,h,l)$ scattering plane at
$(\case{2n+1}{2},\case{2n+1}{2},\case{2m+1}{2})$ ($n$, $m$-
integer) positions. The measured field- and temperature
dependencies of magnetic diffraction intensity at
$\bbox{q}=(0.5,0.5,0.5)$ are shown in Fig.~\ref{vst} (open
symbols). A well-defined transition is observed in agreement with
specific heat measurements (Fig.~\ref{phase}, diamonds).

The apparent Bragg rod intensity for $\bbox{H}||[1,0,0]$ and
$\bbox{H}||[1,-1,0]$ is substantially smaller than the intensity
of magnetic Bragg peaks seen for $\bbox{H}||[0,1,-1]$
(Fig.~\ref{exdata1}). This is due to that fact that in the
2D-ordered phase magnetic intensity is spread out through an
entire Brillouin zone along the $a$ axis, rather than being
concentrated at $h=0$. However, in the 2D phase, within each
$(b,c)$ plane, the actual ordered moment is similar to that in the
3D ordered state. The 2D ordered moment for $\bbox{H}||[1,0,0]$
and $\bbox{H}||[1,-1,0]$ was deduced from an analysis of measured
Bragg rod intensities, paying close attention to resolution
effects associated with vertical angular acceptance of the
spectrometer. A series of rocking scans across the rods were
collected for $H=7$~T$||[1,0,0]$, $T=0.38$~K and
$H=9$~T$||[1,-1,0]$, $T=0.25$~K, respectively. The data were
normalized by the measured nuclear Bragg peak intensities. The
2-dimensional spin structure was assumed to be collinear, with
spins perpendicular to both the chain axis and $\bbox{H}$ in each
case. Nearest-neighbor spins were assumed to be aligned
antiparallel along both the $b$ and $c$ axes. This simple model
was found to agree very well with the available data. The ordered
moment was determined to be $m=0.66(4)$~$\mu_B$ and
$m=1.2(1)$~$\mu_B$ for $H=7$~T$||[1,0,0]$, $T=0.38$~K and
$H=9$~T$||[1,-1,0]$, $T=0.25$~K, respectively. The latter value
agrees very well with the 3D ordered moment measured at
$\bbox{H}||[0,-1,1]$ at the same value of field and temperature.

A careful analysis of the scans across the Bragg rods for
$\bbox{H}||[1,0,0]$ reveals a finite correlation length even
within the quasi-2D ordered $(b,c)$ planes. Along the direction of
strongest coupling, i.e., along the chains, the magnetic peaks are
resolution-limited. This can be seen from Fig.~\ref{exdata1}(e),
where the solid line represents experimental $l$-resolution.
However, along the $b$ axis, where magnetic interactions are
weaker, the scattering rods are visibly broader than experimental
resolution (Fig.~\ref{exdata1}(d), solid line). Fitting the
$k$-scan to a Lorenzian profile convoluted with the resolution
function (Fig.~\ref{exdata1}(d), dashed line) allows us to extract
the intrinsic width of the rod, that corresponds to a real-space
correlation length $\xi_b=100(20)$~\AA$\approx 10 b$. The
high-field phase in this case is thus characterized as a
short-range ordered state with almost-perfect correlations along
the chains and short-range correlations in the transverse
direction.

This type of ``spin freezing'' in NDMAP  is very similar to that
recently found in the $S=1/2$ quasi-1D AF SrCuO$_2$
\cite{Zaliznyak99}. For a 2-dimensional Heisenberg system, the
ground state is disordered and gapless  for half-integer and
odd-integer spin due to ``instanton'' topological defects
\cite{Haldane88}. If magnetic coupling along the third direction
is sufficiently weak, as it is in NDMAP, it will not be able to
restore LRO in the system due to pinning of such defects within
each plane. The similar behavior of two very different systems,
NDMAP and SrCuO$_2$, suggests that anisotropic spin freezing may
be a rather general feature of quasi-low dimensional
antiferromagnets.

What makes the case of NDMAP special is that both true 3D LRO and
spin freezing can occur in the same sample, depending on the
direction of applied field. This may result from the anisotropic
nature of magnetic interactions along the $a$ axis, that, due to
the very long lattice spacing, are expected to be largely dipolar.
The strength and even sign of such coupling depends on the
orientation of the ordered moment, which, in turn, is defined by
the direction of applied field. Changing the field orientation for
NDMAP  is thus a way to tune those magnetic interactions
ultimately responsible for 3D LRO or spin-freezing behavior.

For the basic physics of Haldane spin chains, a very important
experimental result is the observation of a {\it commensurate}
ordering of {\it transverse} (relative to $\bbox{H}$) spin
components above $H_c$. Such staggered LRO in a quasi-1D material
is a direct consequence of a divergence in the transverse spin
correlation function for an isolated chain in an external field at
$q=\pi$. Theory predicts that above $H_c$ a divergence also exists
in the longitudinal correlator, but at a field-dependent
incommensurate wave vector \cite{AffleckH91,Sakai94}. No
incommensurate magnetic peaks were found in our experiments on
NDMAP. Of course, a divergent susceptibility in a single chain
does not guarantee the appearance of LRO at the same wave vector
in a 3D material. It is also conceivable that incommensurate
elastic scattering in NDMAP appears at a different wave vector
transfer perpendicular to the chain axis, due to anisotropy of
inter-chain interactions, and thus escapes detection in the
present study. Further work will be required to fully resolve this
issue.

In summary, our results provide direct experimental evidence of
field-induced commensurate AF order in a Haldane-gap system, and
are in good agreement with theoretical expectations. Specifics of
inter-chain interactions in NDMAP result in a rich phase diagram
with the high-field phase being either a true 3D ordered state or
a highly anisotropic short-range ordered state.

Work at JHU was supported by the NSF through DMR-9801742.  Work at
Brookhaven National Laboratory was carried out under Contract No.
DE-AC02-98CH10886, Division of Material Science, U.S.\ Department
of Energy. This work used instrumentation supported by NIST and
the NSF through DMR-9423101. Work at RIKEN was supported in part
by a Grant-in-Aid for Scientific Research from the Japanese
Ministry of Education, Science, Sports and Culture.

\end{document}